\documentclass[a4,amssymb,amsmath,aps,prb,reprint,superscriptaddress,showpacs]{revtex4-1}
\usepackage{amssymb,amsmath}
\usepackage{color}
\usepackage{xcolor}
\usepackage{graphicx}
\usepackage{tikz}
\usepackage{bm}
\usepackage{hyperref}
\hypersetup{colorlinks=true, citecolor=blue, urlcolor=blue, linkcolor=red}
\usepackage{epstopdf}
\usepackage[space]{grffile}

\newcommand{\cdoso}{Cd$_2$Os$_2$O$_7$}
\newcommand{\naoso}{NaOsO$_3$}

\begin{document}
\title{Spin and orbital excitations through the metal-to-insulator transition in 
Cd$_2$Os$_2$O$_7$ probed with high-resolution RIXS}

\author{J. G. Vale}
\email{j.vale@ucl.ac.uk}
\affiliation{London Centre for Nanotechnology and Department of Physics and Astronomy, University College London (UCL), Gower Street, London, WC1E 6BT, United Kingdom}
\affiliation{Laboratory for Quantum Magnetism, \'Ecole Polytechnique F\'ed\'erale de Lausanne (EPFL), CH-1015, Switzerland}

\author{S. Calder}
\email{caldersa@ornl.gov}
\affiliation{Neutron Scattering Division, Oak Ridge National Laboratory, Oak Ridge, Tennessee 37831, USA}

\author{N. A. Bogdanov}
\altaffiliation{Current address: Max Planck Institute for Solid State Research, Heisenbergstra{\ss}e 1, 70569 Stuttgart, Germany}
\affiliation{Institute for Theoretical Solid State Physics, IFW Dresden, D01171 Dresden, Germany}

\author{C. Donnerer}
\affiliation{London Centre for Nanotechnology and Department of Physics and Astronomy, University College London (UCL), Gower Street, London, WC1E 6BT, United Kingdom}

\author{M. Moretti Sala}
\altaffiliation{Current address: Department of Physics, Politechnico di Milano, Piazza Leonardo da Vinci 32, 20133 Milano, Italy}
\affiliation{European Synchrotron Radiation Facility (ESRF), CS 40220, F-38043 Grenoble Cedex, France}

\author{N. R. Davies}
\affiliation{Clarendon Laboratory, University of Oxford, Parks Road, Oxford, OX1 3PU, United Kingdom}



\author{D. Mandrus}
\affiliation{Materials Science and Engineering, University of Tennessee, Knoxville, TN 37996, USA}
\affiliation{Materials Science \& Technology Division, Oak Ridge National Laboratory, Oak Ridge, TN 37831, USA}

\author{J. van den Brink}
\affiliation{Institute for Theoretical Solid State Physics, IFW Dresden, D01171 Dresden, Germany}

\author{A. D. Christianson}
\affiliation{Neutron Scattering Division, Oak Ridge National Laboratory, Oak Ridge, Tennessee 37831, USA}
\affiliation{Materials Science \& Technology Division, Oak Ridge National Laboratory, Oak Ridge, TN 37831, USA}
\affiliation{Department of Physics and Astronomy, University of Tennessee, Knoxville, TN 37996, USA}

\author{D. F. McMorrow}
\affiliation{London Centre for Nanotechnology and Department of Physics and Astronomy, University College London (UCL), Gower Street, London, WC1E 6BT, United Kingdom}


\begin{abstract}
High-resolution resonant inelastic x-ray scattering (RIXS) measurements ($\Delta E = \text{46~meV}$) have been performed on \cdoso\ through the metal-to-insulator transition (MIT). A magnetic excitation at 125~meV evolves continuously through the MIT, in agreement with recent Raman scattering results, and provides further confirmation for an all-in, all-out magnetic ground state. Asymmetry of this feature is likely a result of coupling between the electronic and magnetic degrees of freedom. 
We also observe a broad continuum of interband excitations centered at 0.3~eV energy loss. This is indicative of significant hybridization between Os $5d$ and O $2p$ states, and concurrent itinerant nature of the system. In turn, this suggests a possible break down of the free-ion model for \cdoso.
\end{abstract}
\maketitle

\clearpage

\section{Introduction}
A number of osmates undergo unconventional metal-to-insulator transitions (MITs) which appear to be driven to some degree by the onset of antiferromagnetic order, notably NaOsO$_3$ and \cdoso. The former has been proposed to be a rare example of a Slater insulator in three dimensions \cite{shi2009, calder2012, du2012, jung2013, lovecchio2013, middey2014, calder2015_naoso3}, albeit with some controversy \cite{kim2016}.
Like \naoso, bulk measurements on \cdoso\ show a direct correspondence between the MIT and onset of antiferromagnetism at $T_{N}=\text{227~K}$ \cite {sleight1974, mandrus2001, yamaura2012, hiroi2015}.
Resistivity data indicate that a BCS-like indirect charge gap ($2\Delta_C=\text{80~meV}$) opens continuously below the MIT.\cite{mandrus2001, hiroi2015}
Optical conductivity measurements reveal similar behavior for the direct optical gap ($2\Delta_O=\text{158~meV}$ \cite{padilla2002, sohn2015}, with the most recent study suggesting a weak decoupling of the two phenomena \cite{sohn2015}. This manifests as the formation of an apparent antiferromagnetic metallic phase between 210~K and $T_{N}$. \footnote{Whether this is representative of the bulk behavior, or an effect limited close to the sample surface, remains an open question.}
These observations, along with density functional theory (DFT) calculations \cite{shinaoka2012}, led to the suggestion that \cdoso\ undergoes a Lifshitz transition (change in topology of the Fermi surface), which is partly driven by the onset of antiferromagnetic order.

Yet there have been comparatively few studies of the magnetic behavior of \cdoso. Yamaura \emph{et al.}~proposed an all-in, all-out (AIAO) $\bm{k}=0$ magnetic structure for \cdoso, based upon resonant elastic x-ray scattering (REXS) data and representation analysis \cite{yamaura2012}. This was confirmed by the authors of the present article using neutron powder diffraction \cite{calder2016_cdoso}. Such a ground state has also been unambiguously found for the isostructural R$_2$Ir$_2$O$_7$ (R = Sm, Nd) \cite{donnerer2016, guo2016}, and is a direct manifestation of the significant spin-orbit coupling (SOC) present in these systems.
Recent measurements have shown that the N\'{e}el temperature $T_N$ is continuously suppressed as a function of applied hydrostatic pressure, going to zero temperature at 36~GPa \cite{wang2018}. Such behavior is consistent with the expectations for a Lifshitz transition and its underlying $T=0$ quantum critical point.

The effect of SOC also extends to the magnetic excitations.
Previous low resolution ($\Delta E=\text{130~meV}$) resonant inelastic x-ray scattering (RIXS) results presented a dispersionless feature in the excitation spectra for \cdoso\ at 170 meV, which could not be explained in terms of intra-t$_{2g}$ spin-flip excitations or a simple spin wave picture \cite{calder2016_cdoso}. Quantum chemistry (QC) calculations using a nearest-neighbor Heisenberg Hamiltonian,\cite{bogdanov2013} which included significant single-ion anisotropy (SIA), and the Dzyaloshinskii-Moriya (DM) interaction:
\begin{equation}\label{ham_Cd227_disp}
\mathcal{H}=J\sum_{ij} \mathbf{S}_i\cdot \mathbf{S}_j + \mathbf{d}\sum_{ij} \mathbf{S}_i\times \mathbf{S}_j + \sum_i \mathbf{S}_i \cdot \mathsf{A} \cdot \mathbf{S}_i,
\end{equation}
had determined the antiferromagnetic exchange parameter $J = 6.3$~meV, considerably smaller than the energy scale of the new peak. 
Based on exact diagonalization (ED) calculations using the same Hamiltonian, it was concluded that the new feature (henceforth referred to as peak A) was a signature of a combination of $\Delta S\!=\!1,2,3$ excitations: 
$S_z = 3/2\rightarrow 1/2$ ($\Delta S_z=1$), $S_z = 3/2\rightarrow -1/2$ ($\Delta S_z=2$), and $S_z = 3/2\rightarrow -3/2$ ($\Delta S_z=3$). The latter of these was found to be the dominant process in the density of states and previously unobserved by RIXS \footnote{Note that the density of states is inequivalent to the RIXS cross-section; the argument is that the single- and bi-magnon processes should be dominant in the RIXS cross-section.}.  
Excitations with $\Delta S=3$ are usually forbidden in the RIXS process; however it was argued that due to strong spin-orbit coupling in the intermediate state, $S$ was no longer a good quantum number. Hence the excited state wavefunction is comprised of a superposition of different spin-orbital states, all of which have a finite overlap with the $S=3/2$ ground state.  
Recent Raman scattering measurements meanwhile reveal two broad features at 130 and 150~meV \cite{nguyen2017}, proposed to arise from $\Delta S=0$ scattering from the two-magnon density of states. It was found that the coupling parameters were generally in good agreement with the previous QC calculations \cite{bogdanov2013}.

\begin{figure*}[t]
\includegraphics{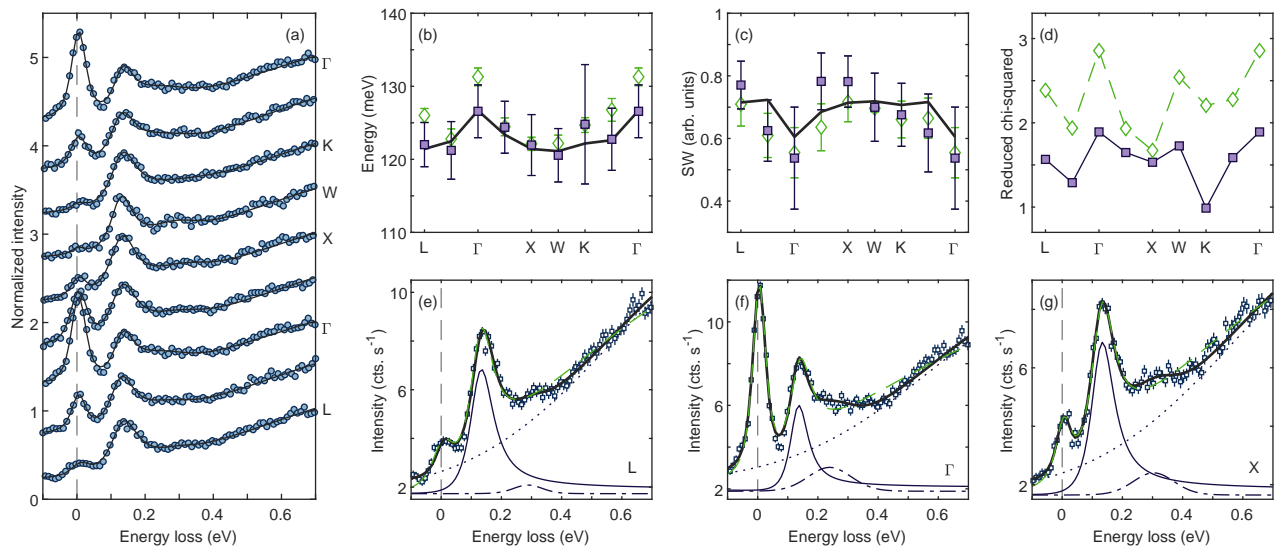}
\caption{(a): RIXS spectra as a function of momentum transfer collected in the $(6,\,7,\,7)$ Brillouin zone at 30~K. Spectra are normalized to the $d-d$ excitations at 0.7~eV and offset for clarity. Black solid lines are the best fit to the model described in main text (including feature at 0.3~eV).
(b--d): Extracted fitting parameters with(out) broad feature at 0.3~eV given by filled squares (open diamonds). The energy and spectral weight of peak A are plotted in (b) and (c), with the corresponding reduced chi-squared of the fits given in (d). Solid line in (b) and (c) is best fit to Hamiltonian given by Eqn.~\ref{ham_Cd227_disp}, with parameters  $J=13.1$~meV, $A=-12.9$~meV, and $\vert\mathbf{d}\vert = 6.8$~meV. 
(e--g): Comparison of the fits at different momentum transfers. Solid and dot-dashed peaks represent Fano resonance and 0.3~eV feature components respectively. Dashed green line indicates fit without 0.3~eV feature. }
\label{Cd227_Qdep_magnon}
\end{figure*}

In this manuscript, we present high resolution RIXS measurements ($\Delta E = \text{46~meV}$) on \cdoso. Our results provide new information about the electronic and magnetic excitations in this material as a function of momentum transfer.
We determine that the magnetic feature (peak A) has a lower energy than previously reported (125~meV), but has minimal dispersion within experimental uncertainty. By modelling the magnetic excitations well below $T_N$, we confirm that peak A includes higher order processes in $\Delta S$, and unambiguously establish that the magnetic ground state is AIAO.
At higher temperatures, the excitation becomes progressively more damped, potentially abating above the MIT. Notably an asymmetry can be observed at all temperatures, which was previously unresolved in the low-resolution RIXS measurements. It is proposed that this results from coupling between the electronic and magnetic degrees of freedom, further highlighting the interplay of these phenomena in this system.
We also find that the orbital excitations exhibit a degree of $5d^4$ character, despite the nominal electronic ground state being $5d^3$. This is likely due to significant hybridization and long-ranged electronic interactions, which go beyond a simple free ion model for \cdoso. 

\section{Experimental details}
RIXS measurements at the Os L$_3$ edge (10.871 keV) were performed on a single crystal of \cdoso\ (ca.~0.5 mm across) on the ID20 spectrometer at the ESRF, Grenoble \cite{moretti2018}. The sample was oriented such that the $\left[1,\,1,\,1\right]$ direction was perpendicular to a copper sample mount, fixed with GE varnish and placed in a closed-flow (Dynaflow) He cryostat. The scattering plane and incident photon polarisation were both horizontal, i.e.~$\pi$--incident polarization, with the incident beam focussed to a size of 20 $\times$ 10 $\mu$m$^2$ (H$\times$V) at the sample position. 

A Si $(6,\,6,\,4)$ four-bounce secondary monochromator was used to define the incident energy, with a Si $(6,\,6,\,4)$ diced spherical analyser (2~m radius, 40~mm diameter mask) used to reflect the scattered photons towards a Medipix CCD detector (pixel size 55~$\mathrm{\mu}$m). Total energy resolution was determined to be $\Delta$E = 46~meV based on scattering from a structural Bragg reflection.

\section{Low temperature behavior}\label{RIXS_lowT}
We first examine the low-energy behavior deep in the antiferromagnetic insulating phase (30~K). This permits detailed study of the spin Hamiltonian towards the localized limit.

\subsection{RIXS}

RIXS spectra collected in the $(6,\,7,\,7)$ Brillouin zone are plotted in Fig.~\ref{Cd227_Qdep_magnon}(a) as a function of momentum transfer. At first glance they appear similar to those presented in Ref.~\onlinecite{calder2016_cdoso}. The main distinction, however, is that peak A exhibits a degree of asymmetry not observed in the previous study.
In order to quantify our observations further, the data were modelled with Gaussians to represent the elastic line and high-energy intra-$t_{2g}$ excitations. A suitable lineshape was added to describe the asymmetric magnetic feature (peak A), and the resultant function was convoluted with the experimental resolution function prior to fitting.
A number of models were considered which take this asymmetry into account, with the best found to be a Breit-Wigner-Fano (BWF) peakshape, given by:
\begin{equation}
I(\omega) = I_0\frac{\left[1+(\omega-\omega_0)/q\Gamma\right]^2}{1+\left[(\omega-\omega_0)/\Gamma\right]^2}. 
\end{equation}
Such a model successfully described the effect of spin-phonon coupling observed in Raman scattering \cite{sohn2017, nguyen2017}. The asymmetry is a direct manifestation of coupling between a resonant mode (with lifetime $\propto 1/\Gamma$, where $\Gamma$ is a damping parameter) and a continuum of states, the magnitude of which is parametrized by the coefficient $|1/q|$.
Note that in the subsequent analysis, it has been assumed that $q\approx 7$ -- indicative of moderate coupling between the resonance and continuum -- is globally fixed for all temperatures and momentum transfers. This is due to insufficient data between the peak maximum and the broad elastic line. In this region the functional form of the lineshape is especially sensitive to the magnitude of $q$.
It was found that the quality of the fit was consistently improved at all temperatures and momentum transfers by the inclusion of a broad, weakly dispersive feature centred at approximately 300~meV [Fig.~\ref{Cd227_Qdep_magnon}(d)]. This feature -- henceforth referred to as peak A2 -- was assumed to have a Gaussian lineshape; potential origins are discussed later.

The energy and spectral weight of peak A are plotted in Fig.~\ref{Cd227_Qdep_magnon}(b,c). 
Within experimental uncertainty, peak A exhibits minimal dispersion throughout the Brillouin zone, with little concurrent weak variation in the spectral weight.
Using the AIAO magnetic ground state and Eqn.~\ref{ham_Cd227_disp} as a starting point, we were able to successfully model the observed behavior within linear spin wave theory (LSWT). This implies that this minimal Hamiltonian is sufficient to describe the magnetic interactions in \cdoso. The coupling parameters we extract, however, are a factor of two larger than determined from quantum chemistry calculations and Raman scattering (Table \ref{Exchange_pars}). This discrepancy is rather puzzling, and we considered alternative explanations. 

\begin{table}[h]
\begin{tabular}{p{4.6cm}|ccccc}
& $J$ & $|\bm{d}|$ & $\mathsf{A}$ & $J/|\bm{d}|$ & $J/\mathsf{A}$\\
\hline
DFT ($U_{\text{eff}}=\text{1.25~eV}$) \cite{shinaoka2012} & 14 & 4 & -24 & 3.5 & -0.6 \\
Quantum chemistry \cite{bogdanov2013} & 6.43 & 1.65 & -6.77 & 3.9 & -1.0 \\
Raman \cite{nguyen2017} (from two-magnon) & 5.1 & 1.7 & -5.3 & 3.0 & -1.0\\
\hline
RIXS (from raw data) & 13.1 & 6.8 & -12.9 & 1.9 & -1.0\\
RIXS (Effective $\Delta S=1$) & 6.5 & 3.4 & -6.5 & 1.9 & -1.0 
\end{tabular}
\caption{Comparison of exchange parameters between different studies.}
\label{Exchange_pars}
\end{table}

\subsection{Verification of magnetic ground state and Hamiltonian}
We first ascertained how the magnitude of different coupling parameters influence the magnetic ground state, and hence, the validity of the putative magnetic Hamiltonian.
An isotropic Heisenberg antiferromagnet on the pyrochlore lattice exhibits geometric frustration. Due to the lattice topology, it is not possible to simultaneously satisfy all of the pairwise interactions between spins. Consequently no long-ranged order is observed down to $T\rightarrow 0$, with the resulting ground state being highly degenerate. 
Introducing anisotropy (or further neighbour interactions) breaks this degeneracy, leading to a unique magnetic ground state.
In the case of zero SIA, the magnetic ground state depends on the sign of the DM vector. If the DM vector is negative (direct DM interaction), then one obtains the AIAO arrangement of spins (irreducible representation $\Gamma_3$). This situation exactly corresponds to that determined for R$_2$Ir$_2$O$_7$ \cite{donnerer2016}.  Meanwhile if the DM vector is positive (indirect DM interaction), then the spins are confined to the $xy$-plane ($\Gamma_5$).
Adding uniaxial SIA along the local $\langle 111\rangle$ direction (compressive trigonal distortion of the OsO$_{\text{6}}$ octahedra) actually helps to stabilize the AIAO ground state. Bogdanov \emph{et al.} argue that this is the dominant driving force for putative AIAO order in \cdoso, with the DM interaction playing a more minor role \cite{bogdanov2013}.

\begin{figure}
\includegraphics[scale=1]{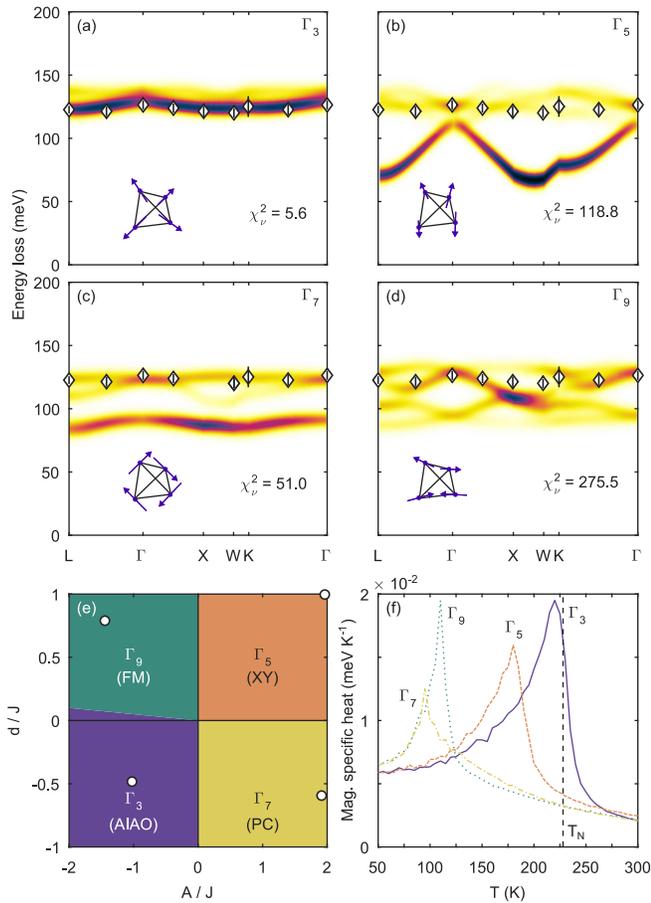}
\caption{(a)--(d): Best fits to the experimental dispersion (open symbols) for different starting magnetic representations $\Gamma_i$. $\Gamma_3$: all-in, all-out (AIAO). $\Gamma_5$: XY. $\Gamma_7$: Palmer-Chalker (PC) phase. $\Gamma_9$: ferromagnet (FM). All figures use the same color scale. (e): Phase diagram reflecting the magnetic ground state for different values of the single-ion anisotropy $A$, and the DM vector $\bm{d} = (d,d,0)$. (f): Magnetic specific heat per moment obtained from simulated annealing ($L=3$, $J=13.1(3)~\text{meV}$). Open symbols in (e) indicate the parameters which were used for the remaining figures.}
\label{phase_diagram}
\end{figure}

The results of a semiclassical energy minimization of a Heisenberg antiferromagnet on the pyrochlore lattice are plotted in Fig.~\ref{phase_diagram}(e) for different values of SIA, and the DM interaction. Note that negative SIA corresponds to a uniaxial anisotropy along the local $\langle 111\rangle$ direction for each moment. 
We find that the resulting ground state spin configuration is highly dependent upon the choice of coupling parameters, with four different ground states occurring in different regions of the phase diagram. These states correspond to the possible magnetic representations $\Gamma_{\text{mag}} = \Gamma_3+\Gamma_5+\Gamma_7+2\Gamma_9$ for $\bm{k}=0$ order on the pyrochlore lattice.
It is possible to go further and calculate the expected spin wave dispersion for a given set of coupling parameters and magnetic structure. In Fig.~\ref{phase_diagram}(a--d), we plot the results of least-squares fits to the experimental dispersion and intensity, which were performed in SpinW \cite{toth2015}. It is clear that an AIAO spin configuration ($\Gamma_3$) provides a much better description of the data for sensible values of the coupling parameters.

Further support for an AIAO ground state can be garnered from results of simulated annealing runs. These were performed on system sizes of $L\times L \times L$ unit cells, where $L$ ranges from 1 to 5 (16 -- 2000 spins in total). The energy of the system and its interactions were determined via the Hamiltonian given by Equation \ref{ham_Cd227_disp}. Coupling parameters used were the same as those obtained from the fits to the spin wave dispersion in Figs.~\ref{phase_diagram}(a)--(d), albeit divided by two \footnote{We provide a natural explanation for this later in the manuscript.}. 
The simulated annealing was performed with a standard Metropolis spin-flip algorithm, with $5\times 10^4$ Monte Carlo steps (MCS) per spin to equilibrate the system for a given temperature, followed by a further $5\times 10^4$ MCS per spin to evaluate thermodynamic parameters.
Two sets of runs were performed for each system size: one with a relatively coarse temperature step ($T_{\text{new}}=0.92\,T$) initialised at 2000 K ($T/J\approx 25$) with the spins oriented randomly, and a dataset focussing on the critical region. The latter was initialised at 320 K (experimentally $T_N + 93$~K) with the state obtained from the coarse dataset at the same temperature, and run with a step-size of 5~K.

The results of these simulations are plotted in Fig.~\ref{phase_diagram}(f) for $L=3$. In each case, the final ground state was the same as that obtained from the semiclassical energy minimization. As expected, the magnetic specific heat $C_m$ diverges at the N\'{e}el temperature, consistent with a second-order phase transition.
For $\Gamma_3$, we find a remarkable agreement between the maximum of $C_m$, and the experimental N\'{e}el temperature ($T_{\text{N}}=\text{227~K}$). The other magnetic representations, however, appear to significantly underestimate $T_{\text{N}}$. 
This suggests that the AIAO spin configuration is indeed the correct one for \cdoso.
\begin{figure}
\centering
\includegraphics[scale=1]{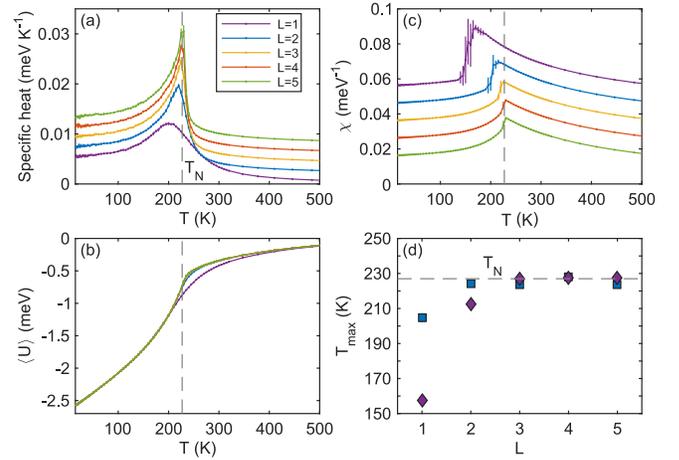}

\caption{Results from simulated annealing runs performed for different system sizes of \cdoso. All data displayed occurs from the mean of five successive runs, with error bars reflecting the standard deviation about this mean. Parameters plotted (per magnetic moment) are the magnetic specific heat $C$ (a), isothermal
susceptibility $\chi$ (b), and mean internal energy $\langle U \rangle$ (c). Curves in (a) and (c) have been offset for clarity.
In (d) the fitted maximum of the specific heat (squares) and susceptibility (diamonds) has been plotted.
Dashed lines indicate the experimental N\'{e}el temperature $T_{\text{N}}$. There appears to be a convergence of the
calculated $T_{\text{N}}$ with the experimental one in the thermodynamic limit.}
\label{Cd227_MC}
\end{figure}
We tested this further by examining the effect of the finite-sized lattice upon the thermodynamic properties (Fig.~\ref{Cd227_MC}). 
The continuous phase transition observed in the magnetic specific heat and susceptibility appears to sharpen and converge towards the experimental N\'{e}el temperature $T_{\text{N}}=227$~K with increasing system sizes. This would be expected as one progresses towards the thermodynamic limit.

There are, however, a number of limitations with the simulations as presented here. Firstly there is a fundamental problem in the Metropolis spin-flip algorithm of `critical slowing-down'. In the vicinity of the critical point, equilibration is slow as both the length and time-scales involved diverge. Thus there are frequently problems with accurate determination of transition temperatures and critical exponents within this approach. Improvements could be made by utilising a cluster-flipping algorithm where critical slowing down is dramatically reduced.
Second, the number of equilibration and averaging steps per spin is relatively low. Ideally one would want to perform the simulations with a greater number of steps to ensure equilibration and accurately determine the thermodynamic parameters with appropriate statistical errorbars. The number of MCS performed per spin was sufficient to ensure equilibration for each temperature and lattice size, however, as determined from the temporal dependence of the internal energy.
Sharp fluctuations can also be observed in the internal energy $\langle U \rangle$ and magnetization at low temperature ($T\!<\!\text{20~K}$). These fluctuations occur over a single temperature step, and may be due to some long timescale behavior which has been incorrectly compensated for in the averaging procedure. Performing a full autocorrelation analysis is likely to improve this.

Finally a complete finite-size scaling analysis has not been performed. Transition temperatures determined from the magnetization, for example, for a given lattice size frequently do not represent the true behavior of the system in the thermodynamic limit. This is why the maxima of the magnetic specific heat and susceptibility curves have been plotted in Fig.~\ref{Cd227_MC}(d) instead of the calculated N\'{e}el temperature. 
Practically what one would do within a finite-size scaling analysis is to calculate the Binder cumulant for each temperature and lattice size, and determine their crossing point via a data collapse. Such an analysis is somewhat subjective, however, since it requires some fine-tuning of the critical temperature and critical exponents to obtain the correct crossing-point.

Despite these limitations, the simulations appear to provide further evidence for an AIAO magnetic ground state in \cdoso. In particular, we obtain remarkable agreement between the calculated and experimental N\'{e}el temperatures for a set of coupling parameters (divided by two relative to the values obtained experimentally) that best fits the observed spin wave dispersion. 
This validates the choice of Hamiltonian, although the factor of two discrepancy with the RIXS results remains puzzling.

\begin{figure*}[t!]
\includegraphics{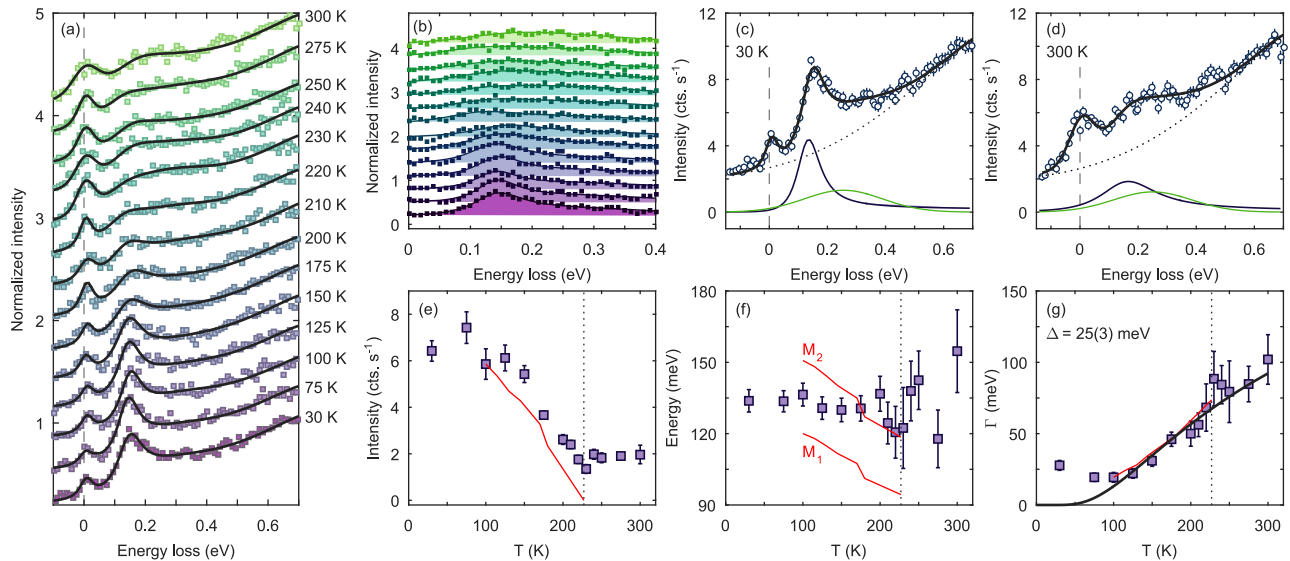}
\caption{Temperature dependence of RIXS spectra collected at $(7,\,7,\,8)$. (a): Stack plot of data normalized to intensity of intra-t$_{\text{2g}}$ excitations at 0.7 eV, plotted with best fit to data. (b): Normalized data with elastic and intra-t$_{\text{2g}}$ contributions subtracted. There is a clear evolution of the lineshape from 30~K (bottom) to 300~K (top) as the gap closes. (c,d): Fits at 30~K (c) and 300~K (d).
(e--g): Temperature dependence of intensity (e), uncoupled resonance energy $\omega_0$ (f), and damping parameter $\Gamma$ (g). Red solid lines are data extracted from Raman scattering measurements \cite{nguyen2017}, which are normalized to 100~K values in (e) and (g). Black line in (g) is best fit to $\Gamma = A\exp{\left(-\Delta/k_B T\right)}$ below $T_{\text{MI}}  $.}
\label{Cd227_Tdep}
\end{figure*}


\section{Temperature dependence}
Now that the low-energy excitations have been characterised deep in the antiferromagnetic insulating phase, we establish how they vary through the MIT. RIXS spectra were collected at the $(7,\,7,\,8)$ Brillouin zone centre $\Gamma$ as a function of temperature. Note that this is in a different Brillouin zone from the momentum dependence data presented in the previous section; the advantage is that the quasielastic peak is noticeably suppressed due to the scattering angle $2\theta$ being closer to the ideal condition of $2\theta=90^{\circ}$.

The results are shown in Fig.~\ref{Cd227_Tdep}. Utilizing the same fitting model as used previously, we observe a clear evolution of the lineshape of peak A from 30~K to 300~K [Figs.~\ref{Cd227_Tdep}(b-d)].
The intensity of this peak decreases continuously as a function of increasing temperature, abating at ca.~230~K [Fig.~\ref{Cd227_Tdep}(e)]. This coincides with the N\'{e}el temperature, and is consistent with the previous RIXS and Raman scattering results \cite{calder2016_cdoso, nguyen2017}.
Meanwhile the peak energy appears approximately constant through the MIT (within experimental resolution) [Fig.~\ref{Cd227_Tdep}(f)]. This contrasts with the Raman scattering data, which shows that the two-magnon peaks $\mathrm{M_1}$ and $\mathrm{M_2}$ are weakly renormalized with temperature. Even so, the general energy scale is consistent. 

Finally the magnitude of the damping parameter $\Gamma$ is approximately resolution-limited at low temperatures and increases rapidly above 150~K [Fig.~\ref{Cd227_Tdep}(g)]. Note that it is unclear within experimental uncertainty whether $\Gamma$ saturates above the MIT, or simply increases monotonically \footnote{The increase in width (and intensity) of the elastic line through the MIT is likely due to low-energy phonon excitations that cannot be discriminated due to the finite energy resolution.}.
There are two possible origins for the observed behavior. The first is that the magnetic excitations become Landau damped by intraband particle-hole excitations. At $T=0$, optical conductivity measurements have determined the optical gap $2\Delta_O = \text{158~meV}$ \cite{sohn2015}. The magnitude of $2\Delta_O$ decreases continuously with increasing temperature, and becomes comparable to the energy of the magnetic excitations at ca.~150~K. Hence one would expect the damping to increase accordingly above this temperature. This scenario is broadly consistent with the data. 
An alternative proposition is that the damping is driven by spin-phonon coupling; as already observed by Raman scattering \cite{nguyen2017} and optical spectroscopy measurements \cite{sohn2017}. From fitting the data above 100~K to an Arrhenius function: $\Gamma = A\exp{\left(-\Delta/k_B T\right)}$, we find that the critical energy scale $\Delta = \text{25(3)~meV}$. This is in good agreement with the lowest energy Os phonon modes, both infrared-active [Os-Os stretch, 24.6(5)~meV] and Raman-active [Os-O(1) stretch, 28~meV]. At present it is not possible to disentangle these two pictures.

\section{Discussion}
\begin{figure}
\includegraphics[scale=1]{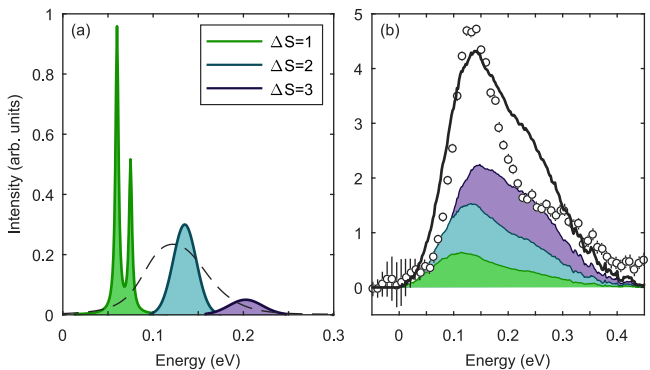}
\caption{Schematic of expected magnetic density of states within linear spin wave theory (a), and that determined within exact diagonalization (b). Dashed line in (a) represents effect of mixing between $\Delta S$ modes. Data points plotted in (b) are the sum of all the data presented in Fig.~\ref{Cd227_Qdep_magnon}(a) with elastic and d-d contributions subtracted off. This is a crude approximation of the RIXS density of states determined from ED.}
\label{mixing}
\end{figure}

Thus far we have determined that the magnetic excitations in \cdoso\ are indicative of coupling between a number of different degrees of freedom. Furthermore, we have unambiguously confirmed the AIAO magnetic ground state and relevant leading order terms in the Hamiltonian.
There are, however, some outstanding questions.
Raman scattering measurements, quantum chemistry calculations, and classical Monte Carlo simulations have all obtained values of the exchange parameters which broadly agree with one another (Table \ref{Exchange_pars}).
Yet the momentum dependence of the feature observed by RIXS appears to be best described by a Hamiltonian dictated by significantly larger coupling parameters ($J=\text{13.1~meV}$, $|\mathbf{d}|=\text{6.8~meV}$, $\mathsf{A}=-\text{12.9~meV}$). We suggest a possible origin for this discrepancy.
In the semiclassical limit (LSWT), and assuming no magnon damping, one would expect magnon branches ($\Delta S=1$) at 65 and 75~meV, and a two-magnon ($\Delta S=0,2$) continuum between 130--150~meV. The kinematic $\Delta S=3$ continuum would be between 195 and 225~meV  [Fig.~\ref{mixing}(a)]. If mixing occurs between these different $\Delta S$ terms, then presumably they would move closer together in energy. The corresponding peak in the imaginary part of the dynamic susceptibility $\chi''(\mathbf{Q}, \omega)$ would consequently be shifted towards the nominal $\Delta S=2$ value. We show this heuristic picture by the dashed line in Fig.~\ref{mixing}(a). Such a scenario would qualitatively agree with the results of the ED calculations as presented in Ref.~\onlinecite{calder2016_cdoso}, which equate to the quantum limit.

Even so, the ED calculations cannot fully describe all of the details observed in the experimental data. 
In order to better compare with theory, all of the data presented in Fig.~\ref{Cd227_Qdep_magnon}(a) have been summed together, and the respective elastic and d-d components subtracted off. This acts as a crude approximation for the magnetic part of the RIXS DOS. We compare this in Fig.~\ref{mixing}(b) with the theoretical RIXS intensity calculated by ED, which is shown \emph{without} the effect of instrumental broadening.
Whilst the energy scale of the excitations agree, there are discrepancies in the lineshape between experiment and theory, especially above 200~meV.

There are a number of factors which are likely to contribute. Firstly, the intensity observed experimentally actually corresponds to the RIXS scattering cross-section, which is significantly more difficult to calculate than the RIXS DOS. It is likely that the RIXS cross-section will be dominated by the $\Delta S=1$ component, as opposed to $\Delta S=3$. Fano coupling will also remove spectral weight at low energies and shift it to higher energies.
Finally the ED calculations used coupling parameters determined from quantum chemistry calculations \cite{bogdanov2013}; Nguyen \emph{et al.} have already shown that the experimental values obtained from Raman scattering appear to be slightly lower \cite{nguyen2017}.

\begin{figure}
\includegraphics{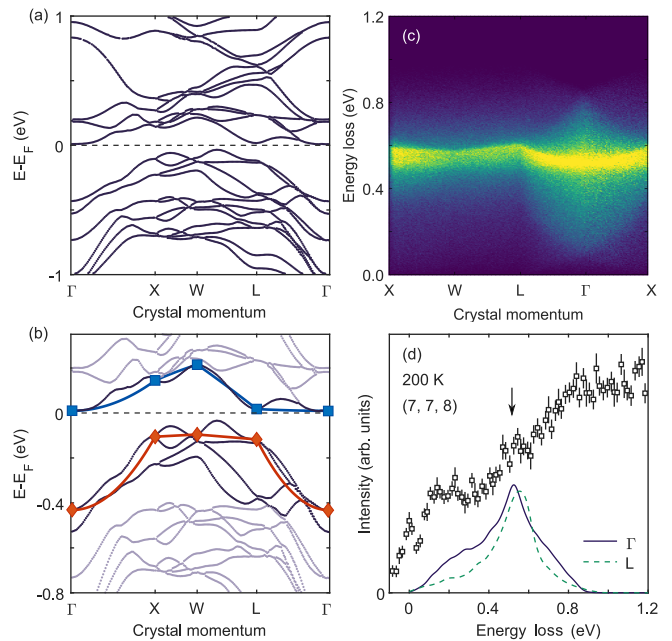}
\caption{(a): Electronic band structure of \cdoso\ close to the Fermi level $E_F$, calculated using density functional theory for $U_{\text{eff}}=\text{1.25~eV}$ \cite{shinaoka2012}. (b): Bands close to $E_F$, along with best fit in the parabolic band approximation. (c): Calculation of dipole-allowed interband transitions between bands close to $E_F$, calculated as described in the main text. (d): RIXS spectrum collected at $(7,\,7,\,8)$ compared with the calculated spectra at $\Gamma$ and $L$. The effect of the instrumental resolution has been included. The arrow highlights a weak feature at 0.6~eV.}
\label{interband}
\end{figure}

Recall that the fit to the data was significantly improved at all temperatures and momentum transfers upon the inclusion of a broad, weakly dispersive feature (peak A2) centered at 0.3~eV energy loss [Fig.~\ref{Cd227_Qdep_magnon}(d)]. No significant temperature dependence was observed [Figs.~\ref{Cd227_Tdep}(c,d)], which implies that it is likely to have an electronic, rather than magnetic, origin.
A similar feature observed in NaOsO$_3$ was suggested to result from inter-band transitions based on calculations of the electronic band structure from density functional theory (DFT) \cite{vale2018_prl, *vale2018_prb}. 
We performed a similar analysis for \cdoso\, using the previously published results by Shinaoka \emph{et al} calculated in the local spin density approximation \cite{shinaoka2012}, including spin-orbit coupling and on-site Coulomb repulsion $U$ ($LSDA+SO+U$) [Fig.~\ref{interband}(a)]. The effective coupling $U_{\text{eff}}=U-J$ was taken to be 1.25~eV, which the authors of Ref.~\onlinecite{shinaoka2012} found gave a ground state which was close to the MIT, but still in the antiferromagnetic insulating phase.
For simplicity, the two electronic bands closest to the Fermi level were fitted in the parabolic approximation, using the band energies at high symmetry points to constrain the fits, and assuming that the band minimum is at $\Gamma$.  
The results are shown in Fig.~\ref{interband}(b), with all the main features of the two electronic bands reasonably reproduced.

Dipole-allowed inter-band transitions were calculated using a Monte Carlo approach. Two wavevectors in the Brillouin zone were randomly selected; these correspond to states in the valence and conduction bands which each have some particular energy. The difference in energy and wavevector between these states was computed, and mapped back onto the first Brillouin zone. This was repeated $4\times10^{8}$ times, with the result shown in Fig.~\ref{interband}(c).
A broad feature can be seen in the data which is centred around 0.55~eV. Along the $\Gamma$-$L$ and $\Gamma$-$X$ directions, however, a continuum of states can be clearly seen which extends to lower energies. 
Comparing the results at $\Gamma$ with experimental data taken at 200~K [Fig.~\ref{interband}(d)], we find a weak peak at 0.6~eV which is consistent with the energy scale of our excitations. Moreover, the low-energy continuum appears to be centered at around 0.25~eV. This suggests that the broad continuum observed previously [Fig.~\ref{Cd227_Tdep}(d)] does indeed manifest from inter-band transitions.

It should be noted that our conclusions come with a number of limitations. First, our calculations only considered the two closest bands to the Fermi level, and assumed they were parabolic. Additional contributions from neighbouring bands will modify the lineshape somewhat, especially at higher energy loss [Fig.~\ref{interband}(a,b)]. The parabolic band approximation also averages over a degree of fine structure which is clearly present from the DFT calculations.
Furthermore, RIXS at the $L_3$ edge is not a direct probe of the band structure, even in weakly correlated systems. Finally, we have neglected likely $\bm{Q}$-dependent matrix element effects, which enter the RIXS cross-section. This may explain the difference in the fitted width of peak A2 for different Brillouin zones at the same high-symmetry point [Figs.~\ref{Cd227_Qdep_magnon}(f), \ref{Cd227_Tdep}(c)].
Nevertheless, the presence of inter-band transitions appears to provide an excellent description of the experimental data.

Our results clearly show the itinerant (non-local) nature of \cdoso. 
Osmium containing transition metal oxides (TMOs) are typically characterized by significant hybridization between the extended 5$d$ orbitals on the osmium site, and the 2$p$ orbitals on the oxygen sites \cite{singh2002, irizawa2006, gangopadhyay2015, taylor2016_sr2scoso6, xu2016, calder2017_ca3lioso6}. For instance, in \naoso, it was determined by DFT that there were on average 4.3 $d$-electrons per Os site, instead of the nominal three \cite{jung2013}.
Quantum chemistry calculations on a small cluster find that the orbital excitations in \cdoso\ are generally better described by a Os $5d^4$ electronic configuration rather than the nominal $5d^3$ (Appendix A). Yet, it has been previously demonstrated that the exchange parameters obtained using a $5d^3$ are in excellent agreement with experiment (Table \ref{Exchange_pars}). 
The complexity of these calculations means that -- by necessity -- they neglect the effect of longer-ranged electronic interactions which may be important. Even so, it highlights that a free-ion description is inadequate for \cdoso.

This is even more apparent when comparing a representative RIXS spectrum of \cdoso\ with that for the nominally isoelectronic (5$d^3$) Ca$_{\text{3}}$LiOsO$_{\text{6}}$ \cite{taylor2017, calder2017_ca3lioso6}. This insulating double perovskite consists of isolated, almost ideal OsO$_{\text{6}}$ octahedra separated by Li cations, and consequently acts as a good approximation to a free-ion model. This is in spite of strong hybridization between Os 5$d$ and O 2$p$ orbitals \cite{calder2017_ca3lioso6}.
Four resolution-limited ($\Delta E=\text{150~meV}$) excitations can be observed for Ca$_3$LiOsO$_6$. This contrasts markedly with the situation for \cdoso, where the features are considerably broader despite the improved resolution ($\Delta E =\text{46~meV}$). 
Note that for $L$-edge RIXS -- in particular on 5d TMOs -- the intrinsic width (in energy) of an arbitrary excitation is essentially proportional to the inverse lifetime of the final state. This contrasts with x-ray absorption spectroscopy (XAS), in which the spectra are broadened by the core-hole in the intermediate state. What we observe experimentally is a convolution of this excitation with the spectrometer resolution function.
Clearly the intrinsic width of these excitations is significantly broader for \cdoso\ than the experimental resolution. This is indicative of the final state having a short lifetime.
We posit this is due to longer-ranged electronic interactions between neighboring OsO$_6$ octahedra through the shared oxygen site (absent in Ca$_3$LiOsO$_6$), which, in turn, leads to greater itinerancy.

\section{Conclusion}
We have presented high-resolution RIXS measurements on \cdoso. Our results highlight the complementary information that RIXS and Raman scattering provide. 
Raman scattering has superior energy resolution and count rates, however it is only sensitive to zone-center $\Delta S=0$ excitations as a consequence of selection rules. Meanwhile RIXS is a momentum resolved, orbital and element-specific technique, which can probe $\Delta S\neq 0$ excitations due to spin-orbit coupling in the intermediate state. 
We conclude that the magnetic excitations previously observed by RIXS and Raman scattering essentially correspond to different parts of the same feature. The asymmetry of peak A is proposed to result from coupling between a resonant magnetic process and inter-band electronic excitations. This further highlights the interplay between magnetism and the electronic behavior of \cdoso. 
Moreover, the presence of inter-band transitions is due to significant hybridization and itinerant behavior, and is indicative of a breakdown of the free-ion model frequently used to describe RIXS spectra. This would have ramifications for the analysis of orbital excitations in other systems proximate to a metal-insulator transition, and ties in with the description of magnetic excitations in the itinerant and localized regimes \cite{vale2018_prl, *vale2018_prb}.

\acknowledgements{
Research in London was supported by the Engineering and Physical Sciences Research Council (Grants No. EP/N027671/1, EP/N034872/1). 
Work at ORNL's High Flux Isotope Reactor was supported by the Scientific User Facilities Division, Office of Basic Energy Sciences, U.S. Department of Energy (DOE). J.G.V. would like to thank University College London (UCL) and \'{E}cole  Polytechnique F\'{e}d\'{e}rale de Lausanne (EPFL) for financial support through a UCL Impact award. A.D.C. was partially supported by the U.S. Department of Energy, Office of Science, Basic Energy Sciences, Materials Sciences and Engineering Division.}

\appendix
\section{Quantum chemistry calculations}

New quantum chemistry calculations were performed for \cdoso\ following the
protocol described in Ref.~\onlinecite{bogdanov2013} for calculating the intra-site excitation energies \cite{molpro, ECP_Stoll_5d, ECP_Dolg_4d, GBas_molpro_2p, ANOs_pierloot_95}.
These calculations simulate the effect of adding one extra electron to the finite cluster, giving rise to a $5d^4$ electronic configuration.
The embedded cluster consists of one reference OsO$_6$ octahedron, six adjacent Cd
sites, and six nearest neighbor OsO$_6$ octahedra in which Os$^{5+}$ were
represented by closed-shell Ta$^{5+}$ $d^0$ species.
The farther environment was modeled as an one-electron effective potential, which
in a ionic picture reproduces the Madelung field in the cluster region.
We performed complete active space self-consistent-field (CASSCF) calculations
with 4 electrons in 5 orbitals averaging over lowest six singlet, ten triplet
and two quintet states.
In the following multireference configuration interaction (MRCI) computation, we
included on top of the CASSCF wave functions all single and double excitations
from the Os $5d$ and O $2p$ orbitals at the central octahedron.
Finally, the SOC Hamiltonian \cite{SOC_molpro} between all spin components of
the 18 spin-free MRCI states is computed and diagonalized resulting in 46 SO
states.
The splittings between the lowest 18 states are listed in Table~\ref{d4_qc}.

We compared the results of these calculations (and those presented in Ref.~\onlinecite{bogdanov2013}, which are reproduced in Table \ref{QC_5d3}) with the experimental data. We find that the agreement with experiment is generally better for a $5d^4$ configuration than $5d^3$. 
There are two main observations which lead to this conclusion.
For instance, previous high-resolution RIXS measurements show a broad excitation centred at 4.5~eV (peak B). This was proposed by the authors to result from multiple overlapping $t_{2g}^3\rightarrow t_{2g}^2e_g$ transitions \cite{calder2016_cdoso}.
Yet the previous QC calculations (Table \ref{QC_5d3}) determine the lowest energy $t_{2g}^2e_g$ multiplet to be at 5.1~eV. This overestimate of the energy scales consistently occurs both within MRCI and MRCI+SOC.
Meanwhile, the energy scale of the $S=1$ $t_{2g}^4\rightarrow t_{2g}^3e_g$ excitations appears to match rather well with the experimental results (Table \ref{d4_qc}). We also note that the MRCI+SOC calculations predict an excitation at 0.36~eV; again comparable to the energy of peak A2.
This suggests that the electronic behavior of \cdoso\ may indeed exhibit some $5d^4$ character.

\begin{table}
\caption{
MRCI and MRCI+SOC relative energies (eV) for the Os $5d^4$ multiplet structure in Cd$_2$Os$_2$O$_7$, given to three significant figures.
The lowest 18 MRCI+SOC states are shown for brevity. Note that the dominant configuration label only corresonds to the states computed using MRCI.}
\label{d4_qc}
\begin{ruledtabular}
\begin{tabular}{l|ll}
dominant config.      &MRCI                       &MRCI+SOC  \\
\hline
$S=1$ $t_{2g}^4$      &0.0; 0.192; 0.196          &0.000; 0.363; 0.364;\\
$S=0$ $t_{2g}^4$      &1.096; 1.097; 1.291;       &0.521; 0.758; 0.776;\\
                      &1.296; 1.322; 2.481        &0.778; 0.800; 0.806;\\
$S=2$ $t_{2g}^3e_g^1$ &2.770; 2.776               &1.74; 1.74; 1.89;\\              
$S=1$ $t_{2g}^3e_g^1$ &4.144; 4.15; 4.514; 		  &1.94; 1.97; 3.19;\\
                      &4.517; 4.520; 4.659;       &3.37; 3.39; 3.40...\\
		 			  &4.683
\end{tabular}						
\end{ruledtabular}
\end{table}
\begin{table}
\caption{
MRCI and MRCI+SOC relative energies (eV) for the Os$^{5+}$ $5d^3$ multiplet structure in Cd$_2$Os$_2$O$_7$. Since cubic symmetry is lifted, the $T$ states are split even without SOC. Each MRCI+SOC value stands for a spin-orbit doublet; for the ${}^4T$ states, only the lowest and highest components are given. Reproduced from Ref.~\onlinecite{bogdanov2013}.}
\label{QC_5d3}
\begin{ruledtabular}
\begin{tabular}{l|ll}
$5d^3$ splittings      &MRCI                       &MRCI+SOC  \\
\hline
${}^4\!A_2 (t_{2g}^3)$       &0.00;                  &0.000; 13.5$\times$10$^{-3}$\\
${}^2\!E (t_{2g}^3)$         &1.51; 1.51             &1.40; 1.53\\
${}^2T_1 (t_{2g}^3)$       &1.61; 1.62; 1.62       &1.63; 1.66; 1.76\\
${}^2T_2 (t_{2g}^3)$       &2.46; 2.49; 2.49       &2.63; 2.76; 2.87\\
${}^4T_2 (t_{2g}^2e_g)$    &5.08; 5.20; 5.20       &5.14; \dots; 5.45\\
${}^4T_1 (t_{2g}^2e_g)$    &5.89; 6.01; 6.01       &6.02; \dots; 6.33\\
${}^4T_1 (t_{2g}^2e_g^2)$  &10.29; 10.63; 10.63    &10.41; \dots; 11.00
\end{tabular}
\end{ruledtabular}
\end{table}

%

\end{document}